\documentclass[preprint,aps,amsmath,amssymb,amsfonts,showpacs,]{revtex4}
\usepackage{epsfig}
\usepackage{mathrsfs}
\usepackage{amsfonts}
\usepackage{amstext}
\usepackage{amsmath}
\usepackage{amssymb}

\def\ra{\rangle}
\def\la{\langle}

\def\be{\begin{equation}}
\def\ee{\end{equation}}
\def\ba{\begin{array}}
\def\ea{\end{array}}

\begin{document}

\fontsize{12pt}{20pt}

\begin{center}\bf  Nonlocality of two-qubit and three-qubit Schmidt-Correlated states \end{center}
\vskip 1mm

\begin{center}
Ming-Jing Zhao$^{1}$, Zong-Guo Li$^{2}$, Bo Li$^{1}$, Shao-Ming Fei$^{1,3}$, Zhi-Xi Wang$^{1}$ and Xian-Qing Li-Jost$^{3}$

\vspace{2ex}

\begin{minipage}{5in}

\small $~^{1}$ {School of Mathematical Sciences, Capital Normal
University, Beijing 100048,
China}

{\small $~^{2}$ College of Science, Tianjin University of Technology, Tianjin,
300191, China}

{\small $~^{3}$ Max-Planck-Institute for Mathematics in the Sciences, 04103
Leipzig, Germany}

\end{minipage}
\end{center}

\vskip 2mm
\parbox{14cm}
{\footnotesize\quad
We investigate the nonlocality %maximal violation of Bell inequalities
of Schmidt-correlated (SC) states, and present analytical
expressions of the maximum violation value of Bell inequalities. It
is shown that the violation of Clauser-Horne-Shimony-Holt (CHSH) inequality is
necessary and sufficient for the nonlocality of two-qubit SC
states, whereas the violation of the Svetlichny inequality is only a
sufficient condition for the genuine nonlocality of three-qubit SC
states. Furthermore, the relations among the maximum violation
values, concurrence and relative entropy entanglement are discussed.}

\bigskip

{{ Keywords}: {\footnotesize  separability; entanglement; Bell
inequality}}

\pacs{03.67.Mn, 03.65.Ud, 03.65.Yz}

{ E-mail address}: zhaomingjingde@126.com (Ming-Jing Zhao)

\maketitle

\section{Introduction}
Einstein, Podolsky, and Rosen (EPR) \cite{epr} believed that the
results of measurements on a local subsystem of a composite physical
system which can be predicted with certainty would be determined by
the local variables of the subsystems. However, the violation of
Bell inequality \cite{bell} rules out all putative local
hidden-variable (LHV) theories, and indicates
%the incompatibility between
%the premises of local realism and the completeness of
%quantum mechanics, and
that quantum nonlocality of entangled states is one of the most profound
characters inherent in quantum mechanics. Moreover, Clauser, Horne, Shimony and Holt
derived the well-known CHSH inequality, which provides a way of
experimental testing of the LHV model \cite{Clauser}.

Actually the nonlocality is intimately related to quantum
entanglement. It is shown that the CHSH inequality is satisfied for
every separable pure two-qubit state, but violated for all entangled
pure two-qubit states, with the amount of violation increasing with
the entanglement \cite{N.Gisin, S.P}. Nevertheless, this conclusion is not true for
mixed entangled states, as
Werner presented a mixed entangled state satisfying the CHSH
inequality \cite{R. F. Werner1989}.
Hence CHSH inequality is just a necessary,
but not sufficient condition for separability of two-qubit states.
Starting with the Bell and CHSH inequalities, many Bell type
inequalities are also proposed with respect to different quantum
systems \cite{bellineqs}. For three-qubit system, Svetlichny
introduced an inequality whose violation is a sufficient
condition for genuine tripartite nonlocality \cite{G. Sve}. Ghose
{\it et al.} further derived the analytical expressions of violation
of Svetlichny inequality for states in Greenberger, Horne and
Zeilinger (GHZ) class
 \cite{S.Ghose}.
However it is still intractable to determine whether a given state, especially mixed state,
violates a certain Bell inequality or not,
as one has to find the mean value of the related Bell
operators for suitable observables \cite{rmp}.

As an important class of mixed states from a quantum dynamical
perspective, % \cite{Khasin06}
Schmidt-correlated (SC) states have
been paid much attention to \cite{Rains,rains1,virmani,ming}.
Just as Khasin {\it et al.} \cite{Khasin06}
proposed, the bipartite SC states naturally appear in a system
dynamics with additive integrals of motion. In
fact, SC states $\rho = \sum_{m,n=0 }^{N-1} a_{mn} |m \cdots m
\rangle \langle n \cdots n|$, $\sum_{m=0}^{N-1} a_{mm}=1$, are
defined as the mixtures of pure states, sharing the same Schmidt
basis \cite{Rains,ming}. The SC states exhibit
some elegant properties. For example, for any local quantum
measurement on SC states, the result does not depend on which party
the measurement is performed. Moreover, their separability is
determined by the positivity of partial transposition \cite{ming}.
%From
%another point of view, SC states are natural generalizations of pure
%state $|\psi\rangle = \alpha |00\rangle +\beta|11\rangle$ in
%two-qubit system and GHZ state $|\phi\rangle = \alpha |000\rangle
%+\beta|111\rangle$ in three-qubit system with
%$|\alpha|^2+|\beta|^2=1$.
%Therefore SC states may
In this paper we investigate the violation of the CHSH inequality
and Svetlichny inequality for SC states. By presenting an analytical
expression of the maximum expectation value $F_{max}$ of CHSH
inequality for two-qubit systems, we show that whether an SC state
violates CHSH inequality is equivalent to whether it is entangled.
For three-qubit systems, we give an analytical expression of the
maximum expectation value $S_{max}$ of the Svetlichny inequality,
and prove that there exist genuine entangled SC states which obey
Svetlichny inequality. Furthermore, the relations between $F_{max}$
and concurrence \cite{woot}, $S_{max}$ and relative entropy
entanglement \cite{relative entropy} for SC states are derived. At
last we illustrate $F_{max}$ and $S_{max}$ are not monotonic under
local operations and classical communications (LOCC) by explicit
examples.

This paper is organized as follows: in section II, we introduce the
CHSH inequality and investigate the maximum expectation value
$F_{max}$ for two-qubit SC states. Then the relation between
$F_{max}$ and concurrence is provided. In Sec. III, the maximum
expectation value $S_{max}$ of the Svetlichny inequality and its
relation to the relative entropy entanglement are studied for
three-qubit SC states. Finally, we conclude with a summary of our
results in Sec. IV.

\section{two-qubit SC states}

The well-known CHSH inequality is shown to be both necessary and
sufficient for the separability of a two-qubit pure state. The
corresponding Bell operator for the CHSH inequality is given by
\begin{eqnarray}
F=AB+ A B^\prime+ A^\prime B- A^\prime B^\prime,
\end{eqnarray}
where the observables $A=\vec{a} \cdot \vec{\sigma}$ and $A^\prime
=\vec{a}^\prime \cdot \vec{\sigma}$ are associated with the first qubit,
$B=\vec{b} \cdot \vec{\sigma}$ and $B^\prime=\vec{b}^\prime \cdot
\vec{\sigma}$ are associated with the second qubit, while $\vec{a}$, $\vec{a}^\prime$, $\vec{b}$
and $\vec{b}^\prime$ are unit vectors, $\vec{\sigma}=(\sigma_x,\sigma_y,\sigma_z)$ with
$\sigma_x$, $\sigma_y$, $\sigma_z$ the Pauli matrices.
$|\la\psi|F|\psi\ra|\leq 2$ holds if and only if the pure state $|\psi\ra$ is separable.

For any mixed two-qubit state $\rho$, the expectation value
$F(\rho)=Tr(\rho\,F)$ satisfies \be\label{2} |F(\rho)|\leq 2 \ee if
$\rho$ admits local hidden variable model. Violation of the
inequality (\ref{2}) implies that the state $\rho$ is entangled. Let
$F_{max}(\rho)=\max_{A, A^\prime, B, B^\prime}F(\rho)$ be the
maximal value of $F(\rho)$ under all possible observables $A$,
$A^\prime$, $B$ and $B^\prime$. One can then decide whether a state
$\rho$ is entangled in terms of the maximum expectation value.

To find the maximum expectation value $F_{max}$ for a given state
$\rho$, we define $\vec{a}=(\sin \theta_a \cos \phi_a, \sin \theta_a
\sin \phi_a, \cos \theta_a)$, and similarly for the unit vectors
$\vec{a}^\prime$, $\vec{b}$ and $\vec{b}^\prime$. In addition, we
define unit vectors $\vec{d}, \vec{d}^\prime$ such that
$\vec{b}+\vec{b}^\prime = 2\vec{d}\cos \phi$ and
$\vec{b}-\vec{b}^\prime=2\vec{d}^\prime\sin \phi$. Thus
\begin{eqnarray}\label{operator d equaltity}
\vec{d} \cdot \vec{d}^\prime \!=\! \cos \theta_d \cos
\theta_{d^\prime} \!+\!\sin \theta_d \sin \theta_{d^\prime} \cos
(\phi_d -\phi_{d^\prime})=0.
\end{eqnarray}
Set $D=\vec{d} \cdot \vec{\sigma}$ and
$D^\prime=\vec{d}^\prime \cdot \vec{\sigma}$, the expectation value
$F(\rho)$ can be written as
\begin{eqnarray}\label{F}
F(\rho)&=&\langle AB \rangle + \langle AB^\prime \rangle +\langle
A^\prime B \rangle - \langle A^\prime B^\prime \rangle
\\\nonumber&=&\langle A(B+B^\prime) \rangle +\langle A^\prime
(B-B^\prime) \rangle  \\\nonumber &=& 2(\langle AD \rangle \cos \phi
+ \langle A^\prime D^\prime \rangle \sin \phi)\\\nonumber &\leq & 2
(\langle AD \rangle^2 + \langle A^\prime D^\prime \rangle^2)^{1/2},
\end{eqnarray}
where we have used the fact that
\begin{eqnarray}\label{x cos}
x \cos \theta + y \sin \theta \leq (x^2 +y^2)^{1/2},
\end{eqnarray}
with the equality holding when $\tan \theta = y/x$.

For a two-qubit SC state $\rho_1$:
\begin{eqnarray*}
\rho_1=a_1 |00\rangle \langle 00| \!+\! a_2 |00\rangle \langle
11|\!+\! a_2^* |11\rangle \langle 00| + a_4 |11\rangle \langle 11|,
\end{eqnarray*}
with $a_1, a_4 \geq 0 $, $a_1+a_4=1$ and $a_1a_4\geq |a_2|^2$. The
first term in Eq. (\ref{F}) with respect to this mixed state $\rho_1
$ turns out to be
\begin{eqnarray}
\label{first term of F rho 1}
\langle AD \rangle &=& \cos \theta_a
\cos \theta_d + 2 ({\rm Re} (a_2) \cos (\phi_a +\phi_d )
-{\rm Im} (a_2) \sin (\phi_a +\phi_d ))\sin \theta_a \sin
\theta_d \nonumber \\
&\leq & \{\cos^2 \theta_d + 4 [{\rm Re} (a_2)
\cos (\phi_a +\phi_d )
-{\rm Im}(a_2)\sin(\phi_a+\phi_d)]^2 \sin^2 \theta_d \}^{1/2}  \nonumber\\
&\leq & \left[ \cos^2
\theta_d + 4 |a_2|^2 \sin^2 \theta_d \right]^{1/2}\nonumber\\
&=& \left[(1-4 |a_2|^2)\cos^2 \theta_d + 4 |a_2|^2 \right]^{1/2},
\end{eqnarray}
where the inequality (\ref{x cos}) has been taken into account. From
Eq. (\ref{F}) and Eq. (\ref{first term of F rho 1}) we have
\begin{eqnarray}
\label{F rho 1}
F(\rho_1) &\!\!\leq\!\!& 2 [(1-4 |a_2|^2)(\cos^2 \theta_d\!+\!
\cos^2 \theta_{d^\prime}) \!+\! 8 |a_2|^2 ]^{1/2} \nonumber \\
&\!\!\leq\!\!& 2[ 1+ 4 |a_2|^2 ]^{1/2}.
\end{eqnarray}
Here we have employed the fact that the maximum of $\cos^2 \theta_d
+ \cos^2 \theta_{d^\prime}$ is 1 according to Eq. (\ref{operator d
equaltity}). The equality in Eq. (\ref{F rho 1}) holds when
$\vec{a}=\vec{z}$, $\vec{a}^\prime= \vec{x}$, $\vec{b}= \sin \phi
\cos \phi_d\, \vec{x} + \sin \phi \sin \phi_d\, \vec{y} + \cos \phi
\,\vec{z}$ and $\vec{b}^\prime=-\sin \phi \cos \phi_d \,\vec{x} -
\sin \phi \sin \phi_d \,\vec{y} + \cos \phi \,\vec{z}$ with $\tan
\phi = 2 |a_2|$ and $\tan \phi_d = -\frac{{\rm Re} (a_2)}{{\rm Im}
(a_2)}$. Therefore, we obtain
\begin{eqnarray}\label{F rho 1 1}
F_{max}(\rho_1)=2\{ 1+ 4 |a_2|^2 \}^{1/2}.
\end{eqnarray}

Furthermore, the maximum expectation value $F_{max}(\rho_1)$ has a
direct relation with its concurrence \cite{woot}, which is an
entanglement measure. The concurrence for a bipartite pure state
$|\psi \rangle$ is defined by $C(|\psi \rangle)= \sqrt{2(1-Tr
\rho_A^2)}$, where the reduced density matrix $\rho_A$ is given by
$\rho_A=Tr_B(|\psi \rangle \langle \psi|)$. The concurrence is then
extended to mixed states $\rho$ by the convex roof, $C(\rho) \equiv
\min _{\{p_i, |\psi _i \rangle \}} \sum_i p_i C(|\psi _i \rangle)$,
for all possible ensemble realizations $\rho= \sum _i p_i |\psi_i
\rangle \langle \psi_i|$, where $p_i \geq 0$ and $\sum_i p_i=1$.
%an entanglement measure
%defined by $N(\rho)=\frac{\|\rho^{PT}\|^2-1}{2}$ \cite{G.Vidal},
%with $\rho^{PT}$ the partial transposition of $\rho$.
For the state
$\rho_1$ one has $C(\rho_1)=2|a_2|$. Hence we get
\begin{eqnarray}\label{2 pure F and N }
F_{max}(\rho_1)=2[1+C^2(\rho_1) ]^{1/2},
\end{eqnarray}
which shows that $F_{max}(\rho_1)$ increases monotonically with $C(\rho_1)$.

The violation of the CHSH inequality has also relations to the dense
coding, which uses previously shared entangled states to send
possibly more information than classical information encoding. The
capacity of dense coding for a given shared bipartite state
$\rho^{AB}$ is given by $\chi=\log_2 d_A +S(\rho_A)-S(\rho)$, with
$S(\rho)=-tr(\rho\log_2 \rho)$ \cite{DC}. $\rho$ is useful for dense
coding if its capacity is larger than $\log_2 d_A$. It is
straightforwardly verified that for two-qubit SC state $\rho_1$,
\begin{eqnarray*}
\chi=&&1-a_1\log_1a_1-a_4\log_1a_4\\
&&+(\frac{1\!\!+\!\!\sqrt{1\!\!-\!\!4a_1a_4\!\!+\!\!4|a_2|^2}}{2}\log_2\frac{1\!\!+\!\!\sqrt{1\!\!-\!\!4a_1a_4\!\!+\!\!4|a_2|^2}}{2}\\
&&+\frac{1\!\!-\!\!\sqrt{1\!\!-\!\!4a_1a_4\!\!+\!\!4|a_2|^2}}{2}\log_2\frac{1\!\!-\!\!\sqrt{1\!\!-\!\!4a_1a_4\!\!+\!\!4|a_2|^2}}{2}),
\end{eqnarray*}
which also increases monotonically with
the maximum expectation value $F_{max}(\rho_1)$ for given $a_1$ and
$a_4$. Hence one has the following equivalent statements for the SC
state $\rho_1$: (i) it is entangled, (ii) it's concurrence is
greater than zero; (iii) it violates CHSH inequality; (iv) it is useful for
dense coding.

Now we generalize two-qubit SC state $\rho_1$ to mixed state
$\rho_2$
\begin{eqnarray}
\label{ge-state1}
\rho_2&& \!=\! b_1|00\rangle \langle 00| + b_2
|01\rangle \langle 01| + b_3 |10\rangle \langle 10|  +b_4 |11\rangle \langle 11| + c_1 |00\rangle \langle 11| +c_1^*
|11\rangle \langle 00|
\end{eqnarray}
with $b_i\geq 0$, $i=1,2,3,4$, $\sum_{i=1}^{4} b_i=1$, $b_1b_4\geq
|c_1|^2$. Nevertheless by similar calculation we can get its maximum
expectation value
\begin{eqnarray}
F_{max}(\rho_2)\!=\!2 \{ (b_1 +b_4-b_2-b_3)^2+4|c_1|^2\}^{1/2},
\end{eqnarray}
which can be obtained by $\vec{a}=\vec{z}$,
$\vec{a}^\prime= \vec{x}$, $\vec{b}= \sin \phi \cos \phi_d \vec{x} +
\sin \phi \sin \phi_d \vec{y} + \cos \phi \vec{z}$ and
$\vec{b}^\prime=-\sin \phi \cos \phi_d \vec{x} - \sin \phi \sin
\phi_d \vec{y} + \cos \phi \vec{z}$ with $\tan \phi = \frac{2
|c_1|}{b_1 +b_4-b_2-b_3}$ and $\tan \phi_d = -\frac{{\rm Re}
(c_1)}{{\rm Im} (c_1)}$.

Although the amount of maximum violation of CHSH inequalities
increases with the entanglement for the SC states, the maximum
expectation value $F_{max}$ is not a legitimate entanglement measure
for two-qubit states, because it does not decrease monotonically
under LOCC. For example, considering a transverse noise channel
\cite{T. Yu} operating on Bell state $|\psi\rangle =
\frac{1}{\sqrt{2}}(|00\rangle +|11\rangle)$, the output state takes
the following form, $\rho_3=\sum_{i,j=1,2}K_i\otimes K_j
|\psi\rangle \langle \psi|K_i^\dagger \otimes K_j^\dagger$, where
the Kraus operators $K_1$ and $K_2$ denote the transverse noise
channel,
\begin{eqnarray}
K_1=\left(
\begin{array}{cc}
\gamma & 0\\
0 & 1
\end{array}
\right),~~~
K_2=\left(
\begin{array}{cc}
0 & 0\\
\omega & 0
\end{array}
\right),
\end{eqnarray}
with time-dependent parameters $\gamma=\exp(-\Gamma t/2), ~~~
\omega=\sqrt{1-\gamma^2}$. By a simplification, the final state,
$\rho_3=\frac{1}{2}[ \gamma^4 |00\rangle \langle 00| + \gamma^2
(|00\rangle \langle 11| +|11\rangle \langle 00|) + (1 +
\omega^4)|11\rangle \langle 11|+ \gamma^2 \omega^2 (|01\rangle
\langle 01| +|10\rangle \langle 10|)]$, is just of the form in Eq.
(\ref{ge-state1}). Therefore the maximum expectation value of
$\rho_3$ is given by
\begin{eqnarray}
\label{F rho 2}
F_{max}(\rho_3)=2\{(2\gamma^4-2\gamma^2+1)^2+\gamma^4\}^{1/2}.
\end{eqnarray}
It is obvious that the maximum expectation value $F_{max}$ is not a
monotonic function of $\gamma$ from Eq. (\ref{F rho 2}). Hence it is
not monotonic with time under LOCC, i.e., $F_{max}$ is not a
legitimate entanglement measure. On the other hand, we can obtain
the concurrence of $\rho_3$, $C(\rho_3)=\gamma^4$, is monotonic with
$\gamma$. For $t>0.265805/\Gamma$, $\rho_3$ does not violate the
CHSH inequality (see FIG. 1). Thus, CHSH inequality can not detect
entanglement of such states, though in fact some of these states are
distillable \cite{Bennett}, as shown in the experimental
demonstration of the ''hidden nonlocality'' in \cite{Kwiat}.
\begin{figure}[!h]
\begin{center}
\scalebox{0.56}[0.5]{\includegraphics {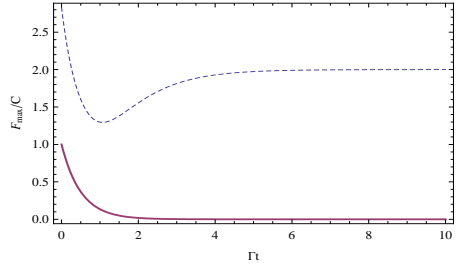}}\label{figure
Ftrho2prime}\caption{Dashed line: $F_{max}(\rho_3)$ versus $\Gamma
t$. Solid line: concurrence $C(\rho_3)$ versus $\Gamma t$.}
\end{center}
\end{figure}

\section{three-qubit SC states}

For three-qubit SC states, we take into account the Svetlichny inequality.
The Svetlichny operator is defined by
\begin{eqnarray*}
S\!&=&\!ABC+ ABC^\prime+ AB^\prime C - AB^\prime
C^\prime + A^\prime BC- A^\prime BC^\prime - A^\prime
B^\prime C- A^\prime B^\prime C^\prime,
\end{eqnarray*}
where observables $A=\vec{a} \cdot \vec{\sigma}$ and $A^\prime
=\vec{a}^\prime \cdot \vec{\sigma}$ are associated with the qubit 1,
$B=\vec{b} \cdot \vec{\sigma}$ and $B^\prime=\vec{b}^\prime \cdot
\vec{\sigma}$ with qubit 2, and $C=\vec{c} \cdot \vec{\sigma}$ and
$C^\prime=\vec{c}^\prime \cdot \vec{\sigma}$ with qubit 3. If a
theory is consistent with a hybrid model of nonlocal-local realism,
then the expectation value for any three-qubit state is bounded by
Svetlichny inequality: $|S(\rho)| \leq 4$, where $S(\rho)=Tr(S\rho)$
is the expectation value of $S$ with respect to state $\rho$. In
this section we are going to derive the analytical expression of
maximum expectation value $S_{max}(\rho) =
\max_{A,A^\prime,B,B^\prime,C,C^\prime} S(\rho)$ for three-qubit SC
states.

In order to find the maximum expectation value $S_{max}$, we
implement the same transformation for $\vec{b}$ and $\vec{b}^\prime$
as in the two-qubit case. The expectation value $S(\rho)$ can be
written as:
\begin{eqnarray}
\label{S}
S(\rho)
&=&\langle ABC \rangle + \langle ABC^\prime \rangle +
\langle AB^\prime C \rangle - \langle AB^\prime C^\prime \rangle
+\langle A^\prime BC \rangle - \langle A^\prime
BC^\prime \rangle -\langle A^\prime B^\prime C \rangle - \langle
A^\prime B^\prime C^\prime\rangle  \nonumber\\
&=& \langle A(B+B^\prime)C \rangle + \langle A(B-B^\prime) C^\prime\rangle
+ \langle A^\prime (B-B^\prime)C \rangle  -
\langle A^\prime (B+B^\prime) C^\prime \rangle\nonumber\\
&=&2(\cos \phi \langle ADC \rangle + \sin \phi \langle AD^\prime
C^\prime \rangle
+ \sin \phi \langle A^\prime D^\prime
C \rangle-\cos\phi\langle A^\prime D C^\prime \rangle) \nonumber\\
&\leq& 2[(\langle ADC \rangle^2 + \langle AD^\prime C^\prime
\rangle^2 )^{1/2}
+( \langle A^\prime D^\prime C
\rangle^2 + \langle A^\prime D C^\prime \rangle^2 )^{1/2}],
\end{eqnarray}
where we have made use of Eq. (\ref{x cos}) again.

For the three-qubit SC state:
\begin{eqnarray*}
\rho_4\!&=&\!a_1 |000\rangle \langle 000|\! + \!a_2 |000\rangle \langle
111| \!
+ \!a_2^* |111\rangle \langle 000| \!+\! a_4 |111\rangle
\langle 111|
\end{eqnarray*}
with $a_1, a_4 \geq 0 $, $a_1+a_4=1$ and $a_1a_4\geq |a_2|^2$. The
first term in Eq. (\ref{S}) with respect to $\rho_4$ is given by
\begin{eqnarray}
\label{first term of S rho 4}
\langle ADC \rangle
&=& (a_1-a_4)\cos
\theta_a \cos \theta_d \cos \theta_c \\\nonumber
&&+ 2 [{\rm Re} (a_2)
\cos(\phi_a+\phi_d+\phi_c)
-{\rm Im} (a_2) \sin
(\phi_a+\phi_d+\phi_c)]\sin\! \theta_a \sin \theta_d \sin \theta_c \nonumber\\
&\!\leq&\! [(a_1\! -\!a_4)^2\!\cos^2 \!\theta_a \cos^2\! \theta_d
\!+\! 4 |a_2|^2 \sin^2 \!\theta_a \sin^2\! \theta_d ]^{\frac{1}{2}}.
\nonumber\\
\end{eqnarray}
From Eq. (\ref{S}) and Eq. (\ref{first term of S rho 4}) we get
\begin{eqnarray}
S(\rho_4) & \leq & 2 \{ [ (a_1-a_4)^2 \cos^2 \theta_a (\cos^2
\theta_d + \cos^2 \theta_{d^\prime})  \nonumber\\ &&+4 |a_2|^2
\sin^2 \theta_a (\sin^2 \theta_d +\sin^2 \theta_{d^\prime}) ]^{1/2}
\nonumber\\ &&+ [(a_1-a_4)^2 \cos^2 \theta_{a^\prime} (\cos^2
\theta_d + \cos^2 \theta_{d^\prime}) \nonumber\\&&+ 4 |a_2|^2 \sin^2
\theta_{a^\prime} (\sin^2 \theta_d +\sin^2 \theta_{d^\prime})
]^{1/2} \}.
\end{eqnarray}
Due to the constraint condition Eq. (\ref{operator d equaltity}),
one has $\cos^2 \theta_d+ \cos^2 \theta_{d^\prime } \leq 1$ and
$\sin^2 \theta_d + \sin^2 \theta_{d^\prime } \leq2$. Therefore we
arrive at
\begin{eqnarray}\label{S rho 4}
S_{max}(\rho_4)= \max \{ 4|1-2a_1|, 8\sqrt{2}|a_2| \}
\end{eqnarray}
from the fact that
\begin{eqnarray}\label{x cos2} x \cos^2 \theta + y
\sin^2 \theta \leq \left\{
   \begin{array}{cc}x, & x \geq y;\\
y, & x\leq y,
   \end{array}\right.
\end{eqnarray}
where the equality holds when $\theta=0$ for the first case, and
when $\theta=\pi/2$ for the second case. Accordingly, $S_{max}(\rho_4)= 4 |1-2a_1|$
holds when $\vec{a},\, \vec{a}^\prime,\, \vec{b},\, \vec{b}^\prime $ are
all aligned along $\vec{z}$, $\vec{c} =sign(1-2a_1)\vec{z}$
%$\vec{c}$ is aligned along $sign(\cos^2
%\theta \gamma^6 +3 \cos^2 \theta \gamma^2 \omega^4 - 3 \cos^2 \theta
%\gamma^4 \omega^2 - \sin^2 \theta - \cos^2 \theta \omega^6) \vec{z}$
and $\vec{c}^\prime =- \vec{c}$, whereas $S_{max}(\rho_4)= 8\sqrt{2}
|a_2| $ holds when all the measurement vectors lie in the $x-y$
plane with $\tan (\phi_a + \phi_d +\phi_c) =\tan( \phi_a
+\phi_{d^\prime} +\phi_{c^\prime} )=\tan( \phi_{a^\prime} +
\phi_{d^\prime} + \phi_c) = -  \frac {{\rm Im}(a_2)}{{\rm
Re}(a_2)}$, $\tan (\phi_{a^\prime} + \phi_{d} + \phi_{c^\prime})
=\pi$, $\phi_d - \phi_{d^\prime} = \frac{\pi}{2}$ and
$\phi=\frac{\pi}{4}$. Eq. (\ref{S rho 4}) implies that $\rho_4$
violates the Svetlichny inequality if and only if $|a_2|>
\frac{1}{2\sqrt{2}}$. However $\rho_4$ is always genuine tripartite
entangled for nonzero $a_2$. Hence the violation of the Svetlichny inequality is only a
sufficient condition for the genuine nonlocality of three-qubit SC
states.
%can not
%detect the genuine tripartite nonlocality of $\rho_4$ for $|a_2|\leq
%\frac{1}{2\sqrt{2}}$.

Now we contrast the violation of Svetlichny inequality with
entanglement. In terms of the reference \cite{lz}, the generalized
concurrence \cite{albev} of three-qubit SC state $\rho_4$
can be obtained, $C(\rho_4)= \sqrt{6}|a_2|$.
Then, the Svetlichny inequality does not hold when $C(\rho_4)\geq \frac{\sqrt{3}}{2}$, and its violation
satisfies the following equation
\begin{eqnarray}
S_{max}(\rho_4)=\frac{8C(\rho_4)}{\sqrt{3}}.
\end{eqnarray}

Moreover, $S_{max}(\rho_4)$ has also direct relations to the
relative entropy entanglement, $ E(\rho) =\min _{ \sigma \in D} S(
\rho
\parallel \sigma ) =\min_{ \sigma \in D} Tr [ \rho \log \rho - \rho \log \sigma ]$,
where $D$ is the set of all fully separable states. It has been proven that $ \varrho =a_{1} |000 \rangle \langle 000 |+a_{4} |111 \rangle \langle 111
|$ is the optimal separable state for $\rho_4$ such that
$E(\rho_4)=\min_{ \sigma \in D}S( \rho_4
\parallel \sigma ) = S( \rho_4 \parallel \varrho )$ \cite{ming}. Hence, when
$\rho_4$ violates Svetlichny inequality, we have
\begin{eqnarray*}
E(\rho_4)\label{relative entropy of rho 4}
% =&&\frac{1+\sqrt{1-4a_1a_4+4|a_2|^2}}{2} \log
%\frac{1+\sqrt{1-4a_1a_4+4|a_2|^2}}{2} \\\nonumber
%&&+\frac{1-\sqrt{1-4a_1a_4+4|a_2|^2}}{2} \log
%\frac{1-\sqrt{1-4a_1a_4+4|a_2|^2}}{2} \nonumber\\
%&&-(\frac{(a_1\log_2a_1 +a_4\log_2a_4)^2+\sqrt{(a_1\log_2a_1
%-a_4\log_2a_4)^2+4|a_2|^2\log_2a_1\log_2a_4}}{2}
%\log_2\frac{(a_1\log_2a_1 +a_4\log_2a_4)^2+\sqrt{(a_1\log_2a_1
%-a_4\log_2a_4)^2+4|a_2|^2\log_2a_1\log_2a_4}}{2}\\\nonumber
%&&+\frac{(a_1\log_2a_1 +a_4\log_2a_4)^2-\sqrt{(a_1\log_2a_1
%-a_4\log_2a_4)^2+4|a_2|^2\log_2a_1\log_2a_4}}{2}
%\log_2\frac{(a_1\log_2a_1 +a_4\log_2a_4)^2-\sqrt{(a_1\log_2a_1
%-a_4\log_2a_4)^2+4|a_2|^2\log_2a_1\log_2a_4}}{2})
&=&f(a_1,a_4,a_2,a_2^*)-f(a_1\log_2a_1,a_4\log_2a_4,a_2\log_2a_4,a_2^*\log_2a_1)\\
&=&g(a_1,a_4,S_{max}^2)-g(a_1\log_2a_1,a_4\log_2a_4,S_{max}^2\log_2a_1\log_2a_4),%\label{25}
\end{eqnarray*}
where $f(x_1,x_2,x_3,x_4)=f_+ \log_2 f_+ +f_- \log_2 f_-$,
$f_{\pm}=[(x_1 +x_2)\pm\sqrt{(x_1-x_2)^2+4x_3x_4}]/{2}$ and
$g(x_1,x_2,x_3)=g_+ \log_2 g_+ +g_- \log_2 g_-$,
$g_{\pm}=[(x_1+x_2)\pm\sqrt{(x_1 -x_2)^2+\frac{x_3}{32}}]/{2}$.
%From Eq. (\ref{relative
%entropy of rho 4}) it can be shown that the relative entropy
%entanglement $E(\rho_4)$ increases monotonically with $S_{max}$ for
%given $a_1$ and $a_4$.

Now we consider the generalization of the three-qubit SC state
$\rho_4$ to mixed state $\rho_5$:
\begin{eqnarray}
\label{rho5}
\rho_5 &=& b_1 |000\rangle \langle 000|+ b_2 |001\rangle \langle
001| + b_3 |010\rangle \langle 010|
+ b_4|100\rangle \langle 100| + b_5|011\rangle\langle 011| +b_6
|101\rangle \langle 101| \nonumber\\
&&+ b_7 |110\rangle \langle 110| +b_8 |111\rangle \langle 111| +c_1
|000\rangle \langle 111|
+ c_1^* |111\rangle \langle 000|.
\end{eqnarray}
%Consider the maximum expectation value of $|S|$ for mixed state
%$\rho_4^\prime $. Since the first term in Eq. (\ref{S}) with respect
%to the mixed state $\rho_4 $ gives
%\begin{eqnarray}
%\langle ADC \rangle &=& (a_1 \gamma^6 + 3a_1 \gamma^2 \omega^4 -
%3a_1 \gamma^4 \omega^2 - a_4 -a_1 \omega^6)\cos \theta_a \cos
%\theta_d \cos \theta_c \\\nonumber  && +2\gamma^3({\rm Re} a_2 \cos
%(\phi_a +\phi_d + \phi_c)- {\rm Im} a_2 \sin (\phi_a +\phi_d +
%\phi_c)) \sin \theta_a \sin \theta_d \sin \theta_c
%\\\nonumber   &\leq& \{ (a_1 \gamma^6 + 3a_1 \gamma^2 \omega^4 -
%3a_1 \gamma^4 \omega^2 - a_4 -a_1 \omega^6)^2 \cos^2 \theta_a \cos^2
%\theta_d + 4 \gamma^6 |a_2|^2 \sin^2 \theta_a \sin^2 \theta_d
%\}^{1/2}
%\end{eqnarray}
%where we have used (\ref{x cos}) with respect to $\theta_c$ and
%$\phi_a +\phi_d + \phi_c$,
For such state, the $S_{max}$ becomes
\begin{eqnarray*}
S_{max}(\rho_5)
=\max\{ 4 |b_1-b_2-b_3-b_4+b_5+b_6+b_7-b_8|,8\sqrt{2} |c_1|\}.
\end{eqnarray*}
Thus $\rho_5$ violates the Svetlichny inequality when $|c_1|
> \frac{1}{2\sqrt{2}}$. Here $S_{max}(\rho_5)= 4 |b_1-b_2-b_3-b_4+b_5+b_6+b_7-b_8|$ holds
when $\vec{a},\, \vec{a}^\prime,\, \vec{b},\, \vec{b}^\prime $ are
all aligned along $\vec{z}$, $\vec{c}
=sign(b_1-b_2-b_3-b_4+b_5+b_6+b_7-b_8)\vec{z}$ and $\vec{c}^\prime
=- \vec{c}$. $S_{max}(\rho_5)= 8\sqrt{2} |c_1|$ holds when all the
measurement directions lie in the $x-y$ plane with $\tan (\phi_a +
\phi_d +\phi_c) =\tan( \phi_a +\phi_{d^\prime} +\phi_{c^\prime}
)=\tan( \phi_{a^\prime} + \phi_{d^\prime} + \phi_c) = -\frac{{\rm
Im}(c_1)}{{\rm Re}(c_1)}$, $\tan (\phi_{a^\prime} + \phi_{d} +
\phi_{c^\prime}) =\pi$, $\phi_d - \phi_{d^\prime} = \frac{\pi}{2}$
and $\phi=\frac{\pi}{4}$.

In particular, let's consider a transverse noise channel operating
on the GHZ state $|\phi\rangle = \frac{1}{\sqrt{2}}(|000\rangle +
|111\rangle)$. Then the final state $\rho_6=\sum_{i,j,l=1,2} K_i
\otimes K_j \otimes K_l |\phi\rangle \langle \phi| K_i^\dagger
\otimes K_j^\dagger \otimes K_l^\dagger=\frac{1}{2}[ \gamma^6
|000\rangle \langle 000| + \gamma^4 \omega^2 (|001\rangle \langle
001| + |010\rangle \langle 010| +|100\rangle \langle 100|) +
\gamma^2 \omega^4 (|011\rangle \langle 011| + |101\rangle \langle
101| +|110\rangle \langle 110|) + (1+ \omega^6) |111\rangle \langle
111| + \gamma^3 (|000\rangle \langle 111| + |111\rangle \langle
000|)]$, which is just of the form in Eq. (\ref{rho5}). Therefore
we have
\begin{eqnarray}
\label{S rho 5 prime}
S_{max}(\rho_6)
&=&\max\{ 2 |\gamma^6 +3 \gamma^2 \omega^4 - 3
\gamma^4 \omega^2 -1- \omega^6|, 4\sqrt{2} \gamma^3\}\nonumber\\
&=&\left\{
   \begin{array}{cc}
   2 (1\!-\!\gamma^6 \!-\!3 \gamma^2 \omega^4 \!+\! 3
\gamma^4 \omega^2 \!+\! \omega^6), & 0\!\leq\! \gamma\! \leq\! \frac{1}{\sqrt{2}};\\
4\sqrt{2} \gamma^3, & \frac{1}{\sqrt{2}}\! \leq\! \gamma \!\leq\! 1,
   \end{array}\right.
\end{eqnarray}
which shows that $\rho_6$ violates the Svetlichny inequality when
$t<0.693147/\Gamma$. Namely the Svetlichny inequality can not detect
the hidden nonlocality any more for $t>0.693147/\Gamma$. From Eq.
(\ref{S rho 5 prime}) and FIG. 2, we can see that $S_{max}(\rho_6)$
is not a monotonic function of time; accordingly we assert that
$S_{max}$ is also not a suitable entanglement measure.

\begin{figure}[!h]
\begin{center}
\scalebox{0.56}[0.5]{\includegraphics
{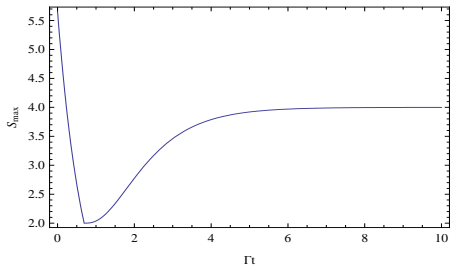}}\caption{$S_{max}(\rho_6)$ versus $\Gamma\, t$}
\end{center}
\end{figure}

\section{Conclusions}

In summary, we have obtained an analytical
formula of maximum expectation value $F_{max}$ of CHSH inequality
for two-qubit SC states, from which we have shown that this inequality is
both necessary and sufficient for the nonlocality of two-qubit SC
states, though  this is not true for general two-qubit mixed
states. In addition, the relations between $F_{max}$, entanglement and capacity
of dense coding for SC states have been also derived. Moreover,
unlike the entanglement measure, $F_{max}$ is not monotonic with
time under LOCC. For three-qubit systems, we have demonstrated that
the violation of the Svetlichny inequality is only a
sufficient condition for the genuine nonlocality of three-qubit SC
states. Furthermore we have
presented a relation between $S_{max}$ and relative entropy
entanglement, which gives a way to determine the relative entropy
entanglement of SC states experimentally.

\section{Acknowledgements}
Ming-Jing Zhao thanks Max-Planck-
Institute for Mathematics in the Sciences for its hospitality.
This work is supported by the NSFC 10875081, NSFC
10871227, KZ200810028013 and PHR201007107 and NSF
of Beijing 1092008.


\begin{thebibliography}{99}

\bibitem{epr}
A. Einstein, B. Podolsky, and N. Rosen, Phys. Rev. \textbf{47}, (1935) 777.

\bibitem{bell}J. S. Bell, Physics (Long Island City, N. Y.) \textbf{1},
(1964) 195.

\bibitem{Clauser} J. F. Clauser, M. A. Horne, A. Shimony, and R. A. Holt, Phys. Rev. Lett. \textbf{23},
(1969) 880.

\bibitem{N.Gisin} N. Gisin, Phys. Lett. A \textbf{154}, (1991) 201.

\bibitem{S.P} S. Popescu and D. Rohrlich, Phys. Lett. A \textbf{166}, (1992) 293.

\bibitem{R. F. Werner1989} R. F. Werner, Phys. Rev. A \textbf{40}, (1989) 4277.

\bibitem{bellineqs}
N. D. Mermin, Phys. Rev. Lett. \textbf{65}, (1990) 1838;\\
M. Ardehali, Phys. Rev. A \textbf{46}, (1992) 5375;\\
N. Gisin and A.
Peres, Phys. Lett. A. \textbf{162},  (1992) 15;\\
R. F. Werner and M.
M. Wolf, Phys. Rev. A \textbf{64}, (2001) 032112;\\
M. \.{Z}ukowski
and \v{C}. Brukner, Phys. Rev. Lett. \textbf{88}, (2002) 210401;\\
M.
\.{Z}ukowski, \v{C}. Brukner, W. Laskowski, and M. Wie\'sniak,
Phys. Rev. Lett. \textbf{88}, (2002) 210402;\\
J. L. Chen, C. F. Wu,
L. C. Kwek, and C. H. Oh, Phys. Rev. Lett. \textbf{93}, (2004) 140407;\\
K. Chen, S. Albeverio, and S. M. Fei, Phys. Rev. A
\textbf{74}, (2006) 050101(R);\\
M. Li and S. M. Fei, Phys. Rev. Lett.
\textbf{104}, (2010) 240502.

\bibitem{G. Sve} G. Svetlichny, Phys. Rev. D \textbf{35}, (1987) 3066.

%\bibitem{J. L. C} J. L. Cereceda, Phys. Rev. A \textbf{66}, 024102 (2002).

\bibitem{S.Ghose} S. Ghose, N. Sinclair, S. Debnath, P. Rungta, and R. Stock, Phys. Rev. Lett. \textbf{102}, (2009) 250404.

\bibitem{rmp} R. Horodecki, P. Horodecki, M. Horodecki, and K.
Horodecki, Rev. Mod. Phys. \textbf{81}, (2009) 865.

\bibitem{Rains}E. M. Rains, Phys. Rev. A \textbf{60}, (1999) 179.

\bibitem{rains1}  E. M. Rains, Phys. Rev. A \textbf{63}, (2000) 019902.

\bibitem{virmani} S. Virmani, M. F. Sacchi, M. B. Plenio, and D. Markham,
Phys. Lett. A \textbf{288}, (2001) 62.

\bibitem{ming} M. J. Zhao, S. M. Fei, and Z. X. Wang, Phys. Lett. A \textbf{372}, (2008) 2552.

\bibitem{Khasin06} M. Khasin and R. Kosloff, Phys. Rev. A \textbf{76}, (2007) 012304.

\bibitem{woot} W. K. Wootters, Phys. Rev. Lett. \textbf{80}, (1998) 2245.

%\bibitem{G.Vidal} G. Vidal and R. F. Werner, Phys. Rev. A \textbf{65}, 032314 (2002).

\bibitem{relative entropy} V. Vedral, M. B. Plenio, M. A. Rippin, and P. L. Knight, Phys. Rev. Lett. \textbf{78}, (1997) 2275;\\
    V. Vedral and M. B. Plenio, Phys. Rev. A \textbf{57}, (1998)
1619.

\bibitem{DC} T. Hiroshima, J. Phys. A \textbf{34}, (2001) 6907;\\
D.
Bru$\rm{\ss}$, G. M. D'Ariano, M. Lewenstein, C. Macchiavello, A.
Sen(De), and U. Sen, Phys. Rev. Lett. \textbf{93}, (2004) 210501.

\bibitem{T. Yu} T. Yu and J. H. Eberly, Phys. Rev. Lett. \textbf{93}, (2004) 140404.

\bibitem{lz} Z. G. Li, in prepration.

\bibitem{albev} S. Albeverio and S. M. Fei, J. Opt. B \textbf{3}, (2001) 223.

\bibitem{Bennett} C. H. Bennett, D. P. DiVincenzo, J. A. Smolin, and W. K. Wootters, Phys. Rev. A \textbf{54}, (1996) 3824 .

\bibitem{Kwiat} P. G. Kwiat, S. Barraza-Lopez, A. Stefanov, and N.
Gisin, Nature, \textbf{409}, (2001) 1014.

\end{thebibliography}
\end{document}